\documentclass[10pt,journal,compsoc]{IEEEtran}
\ifCLASSOPTIONcompsoc
  \usepackage[nocompress]{cite}
\else
  \usepackage{cite}
\fi
\ifCLASSINFOpdf
\else
\fi
\usepackage{amsmath}
\usepackage{url}
\usepackage{graphicx}
\usepackage{multirow}
\usepackage{listings}
\usepackage{algorithm}
\usepackage{algorithmic}
\usepackage{subfigure}
\usepackage[algo2e]{algorithm2e}

\begin{document}
\title{HybridTune: Spatio-temporal Data and Model Driven Performance Diagnosis for Big Data Systems}

\author{Rui~Ren,
        Jiechao~Cheng,
        Xiwen~He,~\IEEEmembership{Student Member,~IEEE,}
        Lei~Wang,Chunjie~Luo,~\IEEEmembership{Member,~IEEE,}
        and~Jianfeng~Zhan,~\IEEEmembership{Senior Member,~IEEE}% <-this % stops a space
%\IEEEcompsocitemizethanks{\IEEEcompsocthanksitem R. Ren, L. Wang and J. Zhan are with the Institute of Computing Technology,
%Chinese Academy of Sciences, No. 6, Kexueyuan South Road, Zhongguancun,
%Haidian District, Beijing, China. \\
%E-mail: {renrui, wanglei_2011, zhanjianfeng}@ict.ac.cn
\IEEEcompsocitemizethanks{\IEEEcompsocthanksitem R. Ren, X. He, L. Wang, C. Luo and J. Zhan are with the Institute of Computing Technology,
Chinese Academy of Sciences, No. 6, Kexueyuan South Road, Zhongguancun,
Haidian District, Beijing, China.
E-mail: {renrui,wanglei\_2011,hexiwen,luochunjie,zhanjianfeng}@ict.ac.cn
\IEEEcompsocthanksitem J. Cheng are with International School of Software, Wuhan University.
E-mail: jetrobert19@gmail.com.}% <-this % stops a space
}


\markboth{Journal of \LaTeX\ Class Files,~Vol.~, No.~, April~2017}%
{Shell \MakeLowercase{\textit{et al.}}: Bare Advanced Demo of IEEEtran.cls for IEEE Computer Society Journals}



\IEEEtitleabstractindextext{%
\begin{abstract}
With tremendous growing interests in Big Data systems, analyzing and facilitating their performance improvement become increasingly important. Although  there have much research efforts for improving Big Data systems performance, efficiently analysing and diagnosing performance bottlenecks over these massively distributed systems remain a major challenge.  In this paper, we propose a spatio-temporal correlation analysis approach based on stage characteristic and distribution characteristic of Big Data applications, which can associate the multi-level performance data fine-grained. On the basis of correlation data,
 we define some priori rules, select features and vectorize the corresponding datasets for different performance bottlenecks, such as, workload imbalance, data skew, abnormal node and outlier metrics. And then, we utilize the data and model driven algorithms for bottlenecks detection and diagnosis. In addition, we design and develop a lightweight, extensible tool HybridTune, and validate the diagnosis effectiveness of our tool with BigDataBench on several benchmark experiments in which the outperform state-of-the-art methods. Our experiments show that the accuracy of abnormal/outlier detection we obtained reaches about 80\%. At last, we report several Spark and Hadoop use cases, which are demonstrated how HybridTune supports users to carry out the performance analysis and diagnosis efficiently on the Spark and Hadoop applications, and our experiences demonstrate HybridTune can help users find the performance bottlenecks and provide optimization recommendations.
\end{abstract}

}

\maketitle

% is not selected <OR> if conference mode is selected - because compsoc
% conference papers position the abstract like regular (non-compsoc)
\IEEEdisplaynontitleabstractindextext

% \fi
\IEEEpeerreviewmaketitle

\ifCLASSOPTIONcompsoc
\IEEEraisesectionheading{\section{Introduction}\label{sec:introduction}}
\else
\section{Introduction}
\label{sec:introduction}
\fi

\IEEEPARstart{R}{ecent} computing industry has witnessed an unprecedented increasing popularity of Big Data. The growth rate of the immense amount of data in our world is doubling faster than ever, and it is estimated that 90\% of the global data has been created in the last two years~\cite{what-is-big-data}. All fields of our lives are now extremely relying on desirable Big Data platforms directly or indirectly, thus Big Data applications like Hadoop~\cite{hadoop}, Spark~\cite{zaharia2012-rdd}, Dryad~\cite{Dryad} and Storm~\cite{Storm} that derive on-going business value became megatrends in enterprises lately.
%Companies, research institutions and governments could personalize their services, optimize decision-making process and predict future trends efficiently and properly through timely and cost-effective analytics over Big Data systems~\cite{Starfish}, owing to the corresponding reduction of operational costs achieved with the power of performance improvements of Big Data systems.
%Big data is rapidly gaining popularity: data volume doubling rates are accelerating, and 90\% of all the data in the world has been created in the past 2 years~\cite{what-is-big-data}. With big data is becoming increasingly important in all areas of our lives, businesses increasingly track and analyze these data in order to find valuable information, and big data system such as Hadoop~\cite{hadoop}, Spark~\cite{zaharia2012-rdd}, Dryad~\cite{Dryad} and Storm~\cite{Storm} are now in widespread use. Timely and cost-effective analytics over Big Data systems has emerged as a key ingredient for success in many businesses, scientific and government endeavors~\cite{Starfish}, because a 10\% improvement in performance of Big Data systems means translating directly to a 10\% reduction in direct operating costs. As a result, both industry and academia have dedicated significant effort towards the performance of big data systems, for example, focusing on a particular aspect of system performance, providing performance analysis tools~\cite{hitune}\cite{SONATA}, and tuning the big data systems~\cite{Starfish}, etc.

\begin{figure*}
 \centering
 \includegraphics[width=7in,height=3in]{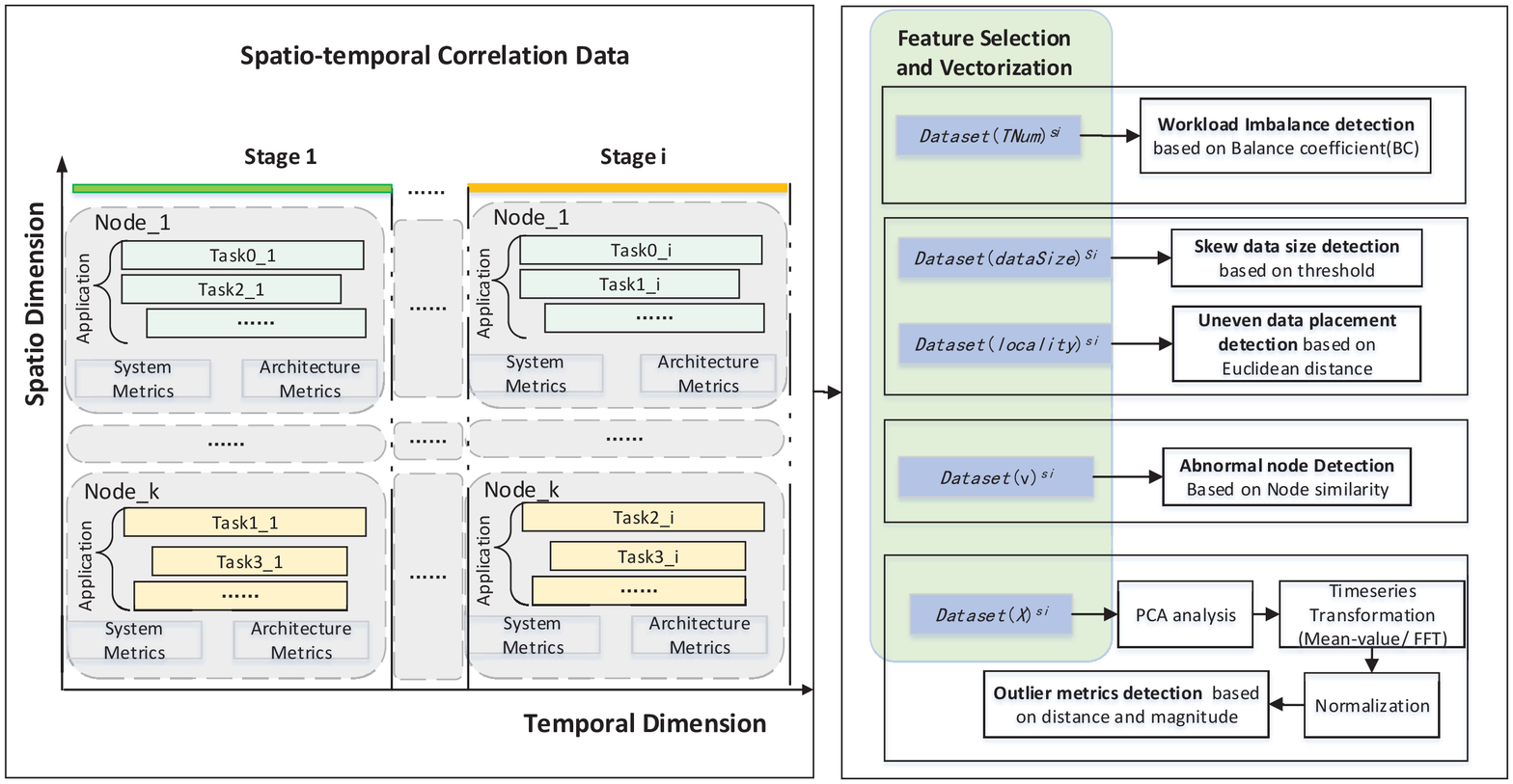}
 \caption{The Performance  Diagnosis Approach based on Spatio-temporal Correlation.}
 \label{analysis_diagnosis}
\end{figure*}

There is no doubt that any level of performance optimization of Big Data systems in the current data-driven society will greatly attract academia and industry concerns.
 However, developing efficient performance analysis and optimization for Big Data systems continues to be a big challenge right now. Because Big Data systems are likely to be constructed from thousands of distributed computing machines, this means that performance issues may exist in a wide variety of subsets or heterogeneous node configurations, such as processors, memory, disks and network; Moreover, the entire software/hardware stacks of Big Data applications are also very complicated and include  hundreds of adjustable parameters, which make performance analysis is more complex and needs fine-grained performance data collection and multi-level data association.
%Currently, there have been a series of works for Big Data systems' performance analysis and optimization.
So far the majority of state-of-the-art performance optimization approaches of Big Data Systems has focused on performance analysis~\cite{hitune}\cite{SONATA}\cite{Theia}\cite{Mochi}\cite{Artemis} and Big Data systems tuning~\cite{Starfish}, etc,. %and much progress has been made in these endeavors.
 %But %For example, most performance analysis tools~\cite{hitune}\cite{SONATA}\cite{Theia} are distributed and scalable, but most tools~\cite{hitune}\cite{SONATA}\cite{Theia}\cite{Artemis} are designed for a particular type of big data system usually and do not consider versatility, which are restrictive when compared to the overall extensibility and versatility that our tool owned.
Although existing studies pay much efforts on overall performance analysis of Big Data Systems in particular, and have made substantial progress.
The limitations of fine-grained locating the root causes of performance bottlenecks with multiple data associated during the life cycle of applications still remain. Moreover, the pure data driven diagnosis approach is promising for relatively simple distributed applications, but it is very time-consuming and difficult to be used in the complex Big Data systems.
And then, we explore and implement HybridTune, which is a hybrid method that combining temporal-spatial data and model. It not only fine-grained diagnose
the performance bottlenecks based on data characteristics, but also significantly reduce the training time based on priori knowledge-based model.
%and expect that our DDTune with multi-level performance data collecting, fine-grained spatio-temporal correlation analysis, and model \& data driven diagnosis.  %will be applicable in various domains.
Concretely, our main contributions are:
\begin{itemize}
 \item We  propose a \emph{spatio-temporal correlation analysis} approach based on stage characteristic and distribution characteristic of Big Data application, which can associate the multi-layer performance data fine-grained. On the basis of correlation data,
      %we also use the \emph{model-based and data-driven} approach for performance bottlenecks detection and diagnosis,
    % we build some suitable datasets for different performance bottlenecks
       we carry out \emph{feature selection and vectorization}, define some prior knowledge, and then use the \emph{data and model driven} approach for performance bottlenecks detection and diagnosis.
     % which can be used for subsequent performance detection and diagnosis based on \emph{model-based and data-driven} algorithms.
\item We design and develop a lightweight, extensible tool HybridTune, with spatio-temporal correlation analysis
and model \& data driven diagnosis approach for Big Data systems.
%It collects the performance data that covering the entire software stack of Big Data platforms (e.g., Hadoop, Spark) effectively, without modifying users program and software.
Then we evaluate diagnosis results of anomaly simulation on distributed systems, and  validate the effectiveness of our tool with BigDataBench on several benchmark experiments in which the outperform state-of-the-art methods. Our experiments show that the accuracy of abnormal/outlier detection we obtained reaches about 80\%.
\item Additionally, we introduce several Spark and Hadoop use cases, and we demonstrate how HybridTune supports users to carry out the performance analysis and diagnosis efficiently on the Spark and Hadoop applications. Above all, our model-based and data-driven detection and  diagnosis methods based on  spatio-temporal correlation data significantly helps to optimize the performance of applications on Big Data platform.
\end{itemize}

The rest of the paper is organized as follows. Section~\ref{background} describes the characteristics of Big Data systems.
%the background of this study.
Section~\ref{diagnosis-method} gives the details of our performance diagnosis methodology. Section~\ref{tool} present the implementation of HybridTune.
Section~\ref{Evaluations}  present experiments and list experimental results. Section~\ref{cases} introduce some case studies. A brief discussion of related works is presented in Section~\ref{related-work}.
%Finally, this work is concluded in Section~\ref{Conclusion}.
Finally, concluding thoughts  are offered in Section~\ref{Conclusion}.

%I wish you the best of success.



%\section{Background}\label{background}
\section{Characteristics of Big Data Systems}\label{background}

The current Big Data systems have two different forms for data processing: (1)batch processing systems, such as  Hadoop\cite{hadoop},  Spark\cite{zaharia2012-rdd}, Dryad\cite{Dryad}, and so on; (2) stream processing system, such as Storm~\cite{Storm} and Spark Streaming~\cite{Sparkstreaming}, etc,. In general, these Big Data systems has two characteristics as follows.

\subsection{Temporal dimension: Stage Characteristics}
%According to the different forms of processing big data, big data systems include batch data processing system  and streaming data processing system. The typical  batch system includes Apache Hadoop \cite{hadoop}, Apache Spark \cite{zaharia2012-rdd}, Dryad \cite{Dryad}, etc, while the typical large-scale streaming system include Storm \cite{Storm} and SparkStreaming \cite{Sparkstreaming}, and so on.

For Big Data applications, we observe that the jobs on different Big Data systems is generally executed in stages or in the stage-like process, and tasks in the same stage are normally executed by the same or similar code on the data partition.
%generally executes in stages or has the execution process like stage, and the tasks of the same stage usually execute the same or similar code on the data partition.
%each job has different stage characteristics for  both batch data processing and streaming data processing using different programming frameworks.
For example, an Hadoop job executes in two stages: Map and Reduce; In the map phase, several map tasks are executed in parallel to process the corresponding input data. After all map tasks finished, the intermediate results are transferred to the reduce tasks for further processing.
While scheduling jobs, Spark splits stages and partition data, then generating the DAG of stage dependency specifically, and the tasks in one stage are executed by the same code and portioned into various data fragments\cite{zaharia2012-rdd}.
%Spark will determine stages and partitions data when scheduling jobs, and generate the DAG of stage dependency; The tasks of one stage execute the same code and process different data fragments.
 %A Dryad job consists of  a set of stages: each stage is composed of an arbitrary number of replicas of a vertex (each operating on a different data partition). The edges of the graph are point-to-point communication channels.
A Dryad job contains a set of stages and each stage consists of an arbitrary set of vertex (each vertex runs on a distinct data partition) replicas, meanwhile all graph edges of the stage constitute point-to-point communication channels~\cite{Dryad}.
 %In addition, the spout and bolt components of Storm also can be seen as the execution stages.
 Similarly, all kinds of spout and bolt components of Storm can also be regarded as the execution stages~\cite{Storm}. In addition,
 Spark Streaming decomposes the batch data into a series of short batch jobs based on Spark~\cite{Sparkstreaming}, so the Spark streaming jobs are executed in stages like Spark jobs.

%Based on this observation, we define the stage characteristics of big data system as \emph{Stage Homogeneous}. By making use of \emph{Stage Homogeneous}, we can detect the performance problems and diagnose causes based on multilevel association analysis for each stage.

\subsection{Spatial dimension: Distribution Characteristics}\
As Big Data systems are used for processing a huge amount of data, a single node is impossible to complete the big tasks in an effective time. So Big Data systems generally use the large-scale distributed cluster architecture. They mainly utilize the distributed file system or distributed database to store data,
 and  use parallel programming model or  distributed execution method  to process/calculate jobs or tasks, that is, the jobs or tasks of Big Data applications will be scheduled into the various nodes in cluster. And the distributed parallel architect distributes data across multiple servers, and it  can turbulence improve data processing speeds. We can consider that, the Big Data applications have distribution characteristics in spatial dimension.
%\section{Methodology} \label{detection-diagnosis}
\section{Performance Diagnosis based on Spatio-temporal Correlation} \label{diagnosis-method}

%\section{Performance Analysis and Diagnosis} \label{detection-diagnosis}

\subsection{Spatio-temporal Correlation }

%According to the temporal and spatial characteristics of Big Data Systems, we propose a \emph{spatio-temporal correlation} approach that involves with timestamp information (e.g., the start time and finish time of stage) and distributed nodes information (e.g., the nodes that tasks run on) to process data in multiple layers.

Generally speaking, jobs on Big Data systems are divided into several stages, and each stage owns multiple tasks, which can run on multiple distributed nodes. According to the temporal and spatial characteristics of Big Data applications, we propose a \emph{spatio-temporal correlation} approach that involves with timestamp information  and distributed nodes information to process the data in multiple layers.

First, in order to implement association in temporal dimension, the time synchronization of  cluster must be guaranteed. Here, the Network Time Protocol (NTP) protocol is used to synchronize computer time in our Big Data systems. Then, combined with the runtime of Big Data applications and the resources utilization (such as, system-level performance metrics and architecture-level performance metrics) in distributed nodes, we classify the correlations between applications and resources into three forms: (1) \emph{task-resource correlation}, (2) \emph{stage-resource correlation} and (3) \emph{job-resource correlation}, according to the execution information (e.g., start time, finish time, nodes of a task, nodes of a stage and nodes of a job).

In our methodology, according to the stage characteristics, we mainly use the \emph{stage-resource correlation} method based on spatio-temporal information  for associating performance metrics between different layers.

\subsection{Feature Selection and Vectorization}

From common sense, we can see that  different performance bottlenecks may lead to different behaviors at the performance data level. So we select the corresponding features and generate vectorization dataset for different performance bottlenecks, such as workload imbalance, data skew, abnormal node and outlier metrics, and the vectorization datasets are as follows:

(1)Workload Imbalance:
We choose the number of tasks on  nodes to describe the workload behavior of applications, and build the dataset $Dataset(TNum)^{s_{i}}$=$\{(node_k,TNum_k)\}^{s_{i}}$. Here, $node_k$ indicates the node $k$, $1 \le k \le p$, and  $p$ is the total number of nodes in cluster; $TNum_k$ indicates the task number on node $k$.

(2)Skew data size:
  We choose the data size of each task processed on each node, and build the dataset $Dataset(dataSize)^{s_{i}}$=$\{(node_j,dataSize_j)\}^{s_{i}}$. Here, $node_j$ indicates to the node with task $j$ execution, $dataSize_j$ indicates to the data size that task $j$ processes.

(3)Uneven data placement:
In our implementation, we utilize the comparison of \emph{data locality} in different tasks or on various nodes to determine whether the uneven data placement exists or not. And we build the dataset $Dataset(locality)^{s_{i}}$=$\{(node_j,locality_j,D^{s_{i}}_{j})\}^{s_{i}}$. Here, $locality_j$  indicates to the \emph{data locality} of task $j$, $D^{s_{i}}_{j}$ indicates to the task runtime of task $j$ in the stage $s_i$ .

(4)Abnormal node:
 we convert collected performance metrics into metrics vector $\overrightarrow{v}$, and generate the metrics vector dataset $Dataset(\overrightarrow{v})^{s_{i}}$=$\{(node_k,\overrightarrow{v_{k}})\}^{s_{i}}$. Here, $\overrightarrow{v_{k}}$=$\{avg(metric_{k1}),...,avg(metric_{kn^*)}\}$ ($avg(metric_{kn^*})$ refers to the average value of $metric_{n^*}$, and $n^*$ refers to the $n^*$-th collected metrics.

(5)Outlier metrics:
 we build the matrix dataset $Dataset(X)^{s_{i}}$=$\{(X)\}^{s_{i}}$, and $X$ is a $m×n^*$ matrix at a stage, columns $n^*$ refers to features number of collected metrics and rows $m$ are collection times during a stage, that is, each row in the matrix determines feature values in a particular timestamp during a stage, for example, $metric_{n^*}^{tm}$ refers to the  $metric_{n^*}$ at the timestamp  $tm$.
$X=\begin{bmatrix} metric_{1}^{t1} & metric_{2}^{t1} & ... & metric_{n^*}^{t1} \\
 metric_{1}^{t2} & metric_{2}^{t2} & ... & metric_{n^*}^{t2}  \\
  ... & ... & ... & ...  \\
metric_{1}^{tm} & metric_{2}^{tm} & ... & metric_{n^*}^{tm}  \end{bmatrix} $

\subsection{Bottlenecks Detection and Diagnosis} \label{diagnostics}

After the multi-level performance data is correlated through the spatio-temporal correlation method,
We propose the automatic bottleneck detection and diagnosis approaches for workload imbalance, data skew, abnormal node and outlier metrics.
For different types of bottlenecks, we first select different features and vectorize these features, and then utilize some model-based and data-driven algorithms for detection or diagnosis.

\subsubsection{Workload imbalance diagnosis}

%If a big data application has uneven distribution of workloads, for example, in the same stage, the task numbers of some nodes are much more or less than other ones, or no task runs on some slave nodes, we consider this  phenomenon as workload imbalance.

 Occasionally, the volume of tasks in certain nodes is bigger or smaller than others at one stage of a Big Data application, or there is no execution  on some slave nodes, we regard this uneven distribution of workloads as workload imbalance.

% \begin{itemize}
%  \item Feature selection and vectorization:  We choose the number of tasks on  nodes to describe the workload behavior of applications, and build the dataset $Dataset(TNum)^{s_{i}}$=$\{(node_k,TNum_k)\}^{s_{i}}$ (here, $node_k$ indicates the node $k$, $1 \le k \le p$, and  $p$ is the total number of nodes in cluster; $TNum_k$ indicates the task number on node $k$ ).
% \end{itemize}

In this section, we use the $Dataset(TNum)^{s_{i}}$ as the input and propose an algorithm to detect whether the task assignment of application is balanced.
Specifically, to quantify \emph{workload imbalance in stage}, we first define  $BC*\overline{TNum^{s_{i}}}$ as the measurement of workload imbalance in stage $s_{i}$. The \emph{balance coefficient ($BC$)} refers to the degree of workload imbalance which can be tolerated, $\overline{TNum^{s_{i}}}$ in Equation~\ref{avgTaskNum} indicates to the average number of tasks in stage $s_{i}$  in the cluster( with $p$ slaves) after removing ultrashort tasks, and $TNum^{s_{i}}_{k}$ refers to the volume of tasks in the stage $s_{i}$ on the node $k$ after removing ultrashort task. Since in some special cases, stages with a number of ultrashort tasks (e.g., failed tasks) running on its nodes greatly affects the judgment of workload imbalance, for that reason we decide to eliminate the ultrashort tasks.
 \begin{equation} \label{avgTaskNum}
 \overline{TNum^{s_{i}}}=\sum TNum^{s_{i}}_{k}/p
\end{equation}

 Then in every stage, in order to determine whether there is workload imbalance in stage, we define $Diff\_TNum_{k}$ to indicate the difference between  $TNum^{s_{i}}_{k}$ and  $\overline{TNum^{s_{i}}}$ on node $k$, if the absolute value $|Diff\_TNum_{k}|$ is greater than the number of tolerated imbalanced tasks $BC*\overline{TNum^{s_{i}}}$, we consider that the workload on this node is imbalanced. Moreover, if the sum $Diff^{s_{i}}$ is greater than $BC*\overline{TNum^{s_{i}}} *p$, we consider there is workload imbalance exist in a stage. In addition, we also use
$Tilt_{k}$ in Equation~\ref{tilt} to  represent the tilt degree of task assignment on node $k$, which is applied to detect the most imbalanced node through ranking methods.
 \begin{equation} \label{tilt}
Tilt_{k}=||Diff\_TNum_{k}|-BC*\overline{TNum^{s_{i}}}|
\end{equation}

 Further more, we assume that the stage number of the job $J$ is $SNum^{s_{i}}$, and  define $Ratio\_UB$ to indicate the ratio of imbalance stage in a job, which is the proportion of the unbalanced stage to the total stage number. If $Ratio\_UB > Th\_UB $ (here, $Th\_UB$ is a threshold of the job having  workload imbalance, which can be set by users), we consider that the job $J$ has workload imbalance. And the Algorithm~\ref{alg-ub} describe how to determine the workload imbalance in stage $s_{i}$ and job $J$.

\begin{algorithm}
\renewcommand{\algorithmicrequire}{ \textbf{Input:}} %Use Input in the format of Algorithm
\renewcommand{\algorithmicensure}{ \textbf{Output:}} %UseOutput in the format of Algorithm
\caption{Algorithm of Determining Workload Imbalance } \label{alg-ub}
\begin{algorithmic}[1]
\REQUIRE ~~\\ %算法的输入参数：Input
$Dataset(TNum)^{s_{i}}$,$BC$,$Th\_UB$,$J$
\ENSURE ~~\\ %算法的输出：Output
 unbalanced $s_{i}$, top $(Tilt\_{k},k)$,$Ratio\_UB$
 \STATE $Ratio\_UB=0 $;
 \STATE $Count\_UB=0$;
  \FOR{$i=1$;$i<n$;$i++$}
   \STATE Set $Diff^{s_{i}}=0$
    \FOR{$k=1$;$k<p$;$k++$}
     \STATE $Diff\_TNum_{k}=TNum^{s_{i}}_{k}-\overline{TNum^{s_{i}}}$;
     \STATE $Diff^{s_{i}}=Diff^{s_{i}}+|Diff\_TNum_{k}|$;
      \STATE  $Tilt_{k}=||Diff\_TNum_{k}|-BC*\overline{TNum^{s_{i}}}|$;
       \STATE Save $Tilt_{k}$ into key-value list $(Tilt_{k},k)$;
    \ENDFOR
     \IF{$Diff^{s_{i}} >  BC*\overline{TNum^{s_{i}}} *p $}
       \STATE  print "Task \ assignment \ at  \ stage \ $s_i$ \ is \  unbalanced";
        \STATE Sort $(Tilt_{k},k)$ from large to small;
        \STATE Output the top $(Tilt_{k},k)$;
        \STATE $Count\_UB++$;
     \ENDIF
  \ENDFOR
  \STATE $Ratio\_UB_=\frac{Count\_UB}{SNum^{s_{i}}}$
  %\STATE $ print " The \ Ratio \ of \ unbalanced \ stage: \ Ratio\_UB " $
  \IF { $Ratio\_UB > Th\_UB$ }
   \STATE Print " Task  assignment  of  job J  is  unbalanced";
  \ENDIF
\end{algorithmic}
\end{algorithm}

\subsubsection{Data skew diagnosis}

Data skew mainly includes two situations: skew data size and uneven data placement.

\paragraph{Skew data size diagnosis}

\begin{algorithm}
\renewcommand{\algorithmicrequire}{ \textbf{Input:}} %Use Input in the format of Algorithm
\renewcommand{\algorithmicensure}{ \textbf{Output:}} %UseOutput in the format of Algorithm
\caption{Skew Data Size Detection Algorithm} \label{size}
\begin{algorithmic}[1]
\REQUIRE ~~\\ %算法的输入参数：Input
$Dataset(dataSize)^{s_{i}}$,$Th\_size$
\ENSURE ~~\\ %算法的输出：Output
The task or node that have skew data size
 \STATE Calculate  $ median(dataSize^{s_{i}})$;
 \STATE Calculate the average value of data size that tasks process on the node k:
 $\overline{dataSize_{k}}$;
 \IF{  $dataSize_{j}/median(dataSize^{s_{i}})>Th\_size$  }
    \STATE Print "the data size of task $j$ is skew";
  \ENDIF
    \IF{ $ \overline{dataSize_{k}}/median(dataSize^{s_{i}})>Th\_size $ }
    \STATE Print "the node $k$ has data size skew";
  \ENDIF
\end{algorithmic}
\end{algorithm}

%For skew data size detection, according to the existing experience, if the data size of a task  processes is much larger than or smaller than other tasks in a stage, we can consider that the task has data skew. If the average data size of a node is much larger than or smaller than other nodes in a stage, we can consider that the node has data skew. If the task or node of a stage has data skew, we can think that the stage has data skew.
From existing experience, data skew phenomenon would exist in a task if the data size of the processing task is much larger or smaller than other tasks in the stage. And similar phenomenon exists in a node if the average size of data is far different from other nodes in a stage, furthermore, data skew of a stage would be deemed to be in existence if data skew was found in the task or node on this stage.

To measure the size of skew data, we use $Dataset(dataSize)^{s_{i}}$=$\{(node_j,dataSize_j)\}^{s_{i}}$ as input, and then we calculate the ratio of data size to the median value of data size  $median(dataSize^{s_{i}})$, afterwards the result of value comparison of the ratio and the threshold $Th\_size$ helps us to measure skew data size. Algorithm~\ref{size} describes the skew data size detection algorithm.

\paragraph{Uneven data placement diagnosis}
\begin{algorithm}
\renewcommand{\algorithmicrequire}{ \textbf{Input:}} %Use Input in the format of Algorithm
\renewcommand{\algorithmicensure}{ \textbf{Output:}} %UseOutput in the format of Algorithm
\caption{Uneven Data Placement Detection based on Euclidean Distance Outlier Algorithm} \label{locality}
\begin{algorithmic}[1]
\REQUIRE ~~\\ %算法的输入参数：Input
$Dataset(locality)^{s_{i}}$
\ENSURE ~~\\ %算法的输出：Output
$Ratio(locality^l,k)$
 \STATE Calculate $ median(D^{s_{i}}) $, standard deviation $std(D^{s_{i}})$;
 \STATE Calculate the distance $dis$ from each $D^{s_{i}}_{j}$ to $ median(D^{s_{i}})$:
 $dis_j=D^{s_{i}}_{j}-median(D^s_{i})$;
   \FOR{$ j=1;j<Num^{s_{i}};j++$}
  %\FOR{ each $D^{s_{i}}_{j}$ in input dataSet:}
   \STATE Summarize: $ sum(dis) = \sum(dis_j)$;
     \STATE Calculate the mean value of $dis$: $mean(dis)$;
      \IF{$|dis_j| > mean(dis)$}
       \STATE Put $D^{s_{i}}_{j}$ into the suspicion group $SuspG$;
      \ENDIF
   \ENDFOR
  \FOR{each $D^{s_{i}}_{j}$ in $SuspG$}
   \IF{  $||dis_j| - mean(dis)| > std(D^{s_{i}}) * 1.96 $   }
     \STATE  Put this $D^{s_{i}}_{j}$ into $outlier\_list(D)$;
     \STATE  Find the corresponding $node_j$ and $locality_j$ of this $D^{s_{i}}_{j}$;
   \ENDIF
   \ENDFOR
  \STATE $Ratio\_UP$=0;
    \FOR{ $k=1;k<p;k++$}
      \FOR{ each $locality^l$ in locality categories}
      \STATE Calculate $Num(oulierD)_k^{locality^l}$  ;
         \STATE Calculate  $Ratio(locality^l,k)$;
         \IF{$Ratio(locality^l,k) > 0$ }
           \STATE  Output $Ratio(locality^l,k)$,and its corresponding node $k$ and $locality^l$, which has uneven data placement.
           \ENDIF
        % \STATE Calculate $Ratio\_UP=Ratio\_UP+Ratio\_locality^l$;
         \ENDFOR
         \ENDFOR
\end{algorithmic}
\end{algorithm}

  Data placement is another critical factor for task runtime and workload imbalance. Because the cluster hardware and workloads are different in the distributed cluster environment, the partitioned data may be placed unevenly, and the task runtime can be very different. In order to find uneven data placement, we focus on the \emph{data locality} of Big Data systems.
 %, which indicates the spatial closeness of data and code.
 %Specifically, Spark's priority of data locality  includes: (1)PROCESS\underline{ }LOCAL, data and code are at the same JVM, (2)NODE\underline{ }LOCAL, data and code are at the same node, (3)NO\underline{ }PREF, there is no difference when the data is processed in any place, which means it has no locality preference, (4)RACK\underline{ }LOCAL, data and code are at the same rack, (5)ANY, data and code are at different machine cross racks. From PROCESS\underline{ }LOCAL to  ANY, they are ordered from the highest priority to the lowest. And Hadoop's data locality includes: (1)NODE\underline{ }LOCALITY, (2)RACK\underline{ }LOCALITY and (3)OFF\underline{ }SWITCH, which is also ordered from the highest priority to the lowest.
  %So we can compare the data locality of different tasks or nodes to judge the uneven data placement.

  %\begin{itemize}
 %\item Feature selection and vectorization:  In our implementation, we utilize the comparison of \emph{data locality} in different tasks or on various nodes to determine whether the uneven data placement exists or not. And we build the dataset $Dataset(locality)^{s_{i}}$=$\{(node_j,locality_j,D^{s_{i}}_{j})\}^{s_{i}}$ (here, $locality_j$  indicates to the \emph{data locality} of task $j$, $D^{s_{i}}_{j}$ indicates to the task runtime of task $j$ in the stage $s_i$ ).
%\end{itemize}

 At the very beginning, we obtain $Dataset(locality)^{s_{i}}$, and then we classify the runtime of localities into two categories by outlier detection algorithm of Euclidean distance, definitely the abnormal runtime ($oulierD^{s_{i}}_{j}$) is much longer than the normal runtime.
Next, we determine whether the data placement leads to abnormal runtime or not, by setting several different weights for distinct localities given the priority of locality. $locality^l$ and $pri(locality^l)$ are utilized here to infer to categories of locality and weights of locality respectively,
for example, we set $pri(RACK\_LOCAL/NODE\_LOCALITY)$=1, $pri(ANY/OFF\_SWITCH)$=2, and some weights are set as 0 ( 0 means these localities are not supposed to cause uneven problems of the data placement).
 Meanwhile, we calculate $Num(oulierD)_k^{locality^l}$, which refers to the number of abnormal runtime of $locality^l$ on node $k$. In addition, we  define  $Ratio(locality^l,k)$ in Equation~\ref{ratio-locality}, and it indicates to the ratio of uneven data placement on a node. When $Ratio(locality^l,k)$ is bigger than 0, the uneven data placement occurs. The uneven data placement detection algorithm is based on  Euclidean distance, and the the whole procedure is demonstrated in Algorithm~\ref{locality}.
\begin{equation} \label{ratio-locality}
\begin{split}
Ratio(locality^l,k)= \frac{Num(oulierD)_k^{locality^l}}{Num^{s_{i}}}* pri(locality^l)
%Ratio\_oulierD^{s_{i}}_{j}* pri(locality^l)   \\
\end{split}
\end{equation}

\subsubsection{Abnormal node diagnosis}\label{abnormal_node_detection}

Execution behaviors of various tasks at the same stage show striking similarity while these tasks running on the homogeneous cluster, thus characteristics of nodes in the homogeneous cluster at one stage are supposed to be analogous. When a node shows significantly different characteristics compared to other nodes at the same stage, the node would be regarded as an \emph{abnormal node} with the potential bottleneck. In this section, we figure out \emph{cosine similarity} between nodes to check abnormal machines.

%In order to find abnormal machines, we compute node similarity based on stage characteristic. Because execution behaviors of tasks at the same stage is similar when tasks runs on the homogeneous cluster, we suppose that node features of homogeneous cluster are similar during the same stage. If one node's feature is significantly different from the other ones during the same stage, we infer that this node is an \emph{abnormal node}, which is a potential bottleneck node.

%\begin{itemize}
 %\item Feature selection and vectorization: we convert collected performance metrics into metrics vector $\overrightarrow{v}$, and generate the metrics vector dataset $Dataset(\overrightarrow{v})^{s_{i}}$=$\{(node_k,\overrightarrow{v_{k}})\}^{s_{i}}$, here $\overrightarrow{v_{k}}$=$\{avg(metric_{k1}),...,avg(metric_{kn^*)}\}$ ($avg(metric_{kn^*})$ refers to the average value of $metric_{n^*}$, and $n^*$ refers to the $n^*$-th collected metrics).
%\end{itemize}

%To check abnormal machines, we figure out \emph{cosine similarity} between nodes.
First of all, we obtain $Dataset(\overrightarrow{v})^{s_{i}}$,
%convert collected performance metrics into metrics vector $\overrightarrow{v}$,
and then we calculate the cosine similarity between the metrics vector on node k ($\overrightarrow{v_{k}}$) and node $k^*$ ($\overrightarrow{v_{k^*}}$), as can be seen in Equation~\ref{cosine}. The closer the cosine value is to 1, the smaller angle between two vectors and the more similar nodes we get.
\begin{equation} \label{cosine}
simi(\overrightarrow{v_{k}},\overrightarrow{v_{k^*}})=\cos { (\theta )}
=\frac{\overrightarrow{v_{k}}\cdot\overrightarrow{v_{k^*}}}{||\overrightarrow{v_{k}}||*||\overrightarrow{v_{k^*}}||}
\end{equation}

 For the sake of detecting the abnormal node, we abandon pairwise comparison method that lacks intuitive results, instead, we measure the average similarity $simi(\overrightarrow{v_{k}},\overrightarrow{v_{others}})$ between each node and all rest nodes shown in Equation~\ref{AvgSimilarity}.
 If $simi(\overrightarrow{v_{k}},\overrightarrow{v_{others}})$ of node $k$ is smaller than a specified similarity threshold $Th\_simi$, then the node $k$ is regarded as an \emph{abnormal node}. Here, $\{Slaves\setminus k\}$  refers to slave nodes set without node $k$.
\begin{equation} \label{AvgSimilarity}
simi(\overrightarrow{v_{k}},\overrightarrow{v_{others}})=\frac{\sum_{node\in \{Slaves \setminus k\} }simi(\overrightarrow{v_{k}},\overrightarrow{v_{node}}) }{p-1}
\end{equation}

\subsubsection{Outlier metrics diagnosis}

Generally, if there are  abnormal nodes during a stage, at the micro level, by observing from micro-level, individual  metrics of abnormal nodes  always have abnormal states; Even if some nodes only subject to interferences, the interfered metrics will also behave differently; So the metrics whose behaviors have a greater difference between  the metrics's  behaviors on other nodes, which are regarded as outlier metrics. In this section, we compare the differences between the principal component metrics at each node in the cluster, and try to find the root cause of performance bottlenecks by observing  the anomalies of metrics.

 %In order to find the reasons of node dissimilarity, we use two methods to comprehensively determine the \emph{critical performance metrics} that affect performance.

\paragraph{Principal Component Analysis}

According to the observations, we learn that not all performance metrics are closely associated with performance anomalies, for some metrics remain stable even an outlier appears. Furthermore, different applications and stages are sensitive to different metrics, hence we use principal component analysis (PCA) ~\cite{pca} for relevant metrics selection.
In general terms, PCA uses an orthogonal transformation to convert to a large set of data observations. The number of principal components is usually less than or equal to original variables, and the first principal component accounts for the most variability in the data. Before using PCA, we first construct the dataset $Dataset(X)^{s_{i}}$. For the matrix $X$, the principal components $\{c_w\}_{w=1}^{n^*}$ can be obtained from Equation ${c_w=argmax_{||x||=1}||(X-\sum_{s=1}^{w-1} X{c_s}{c_s}^T)x||}$, they are the $n^*$ eigenvectors of the covariance matrix $CM=\frac{1}{m}X^TX$~\cite{Eagle}.
 % \begin{itemize}
% \item Feature selection and vectorization: we build the matrix dataset $Dataset(X)^{s_{i}}$=$\{(X)\}^{s_{i}}$, and $X$ is a $m×n^*$ matrix at a stage, columns $n^*$ refers to features number of collected metrics and rows $m$ are collection times during a stage (that is, each row in the matrix determines feature values in a particular timestamp during a stage, for example, $metric_{n^*}^{tm}$ refers to the  $metric_{n^*}$ at the timestamp  $tm$).
%\end{itemize}
%$X=\begin{bmatrix} metric_{1}^{t1} & metric_{2}^{t1} & ... & metric_{n^*}^{t1} \\
 %metric_{1}^{t2} & metric_{2}^{t2} & ... & metric_{n^*}^{t2}  \\
%  ... & ... & ... & ...
 % \\ metric_{1}^{tm} & metric_{2}^{tm} & ... & metric_{n^*}^{tm}  \end{bmatrix} $

 In addition, the cumulative contribution rate of PCA is used in our evaluation for selecting and determining an appropriate dimension of eigenspace. The rate formula is $CCRate_d=\frac{\sum_{w=1}^{d}c_w}{\sum_{w=1}^{n^*}c_w}$, and $d$ represents the top $d$ principal components.

\paragraph{Time Series Transformation}

%We first define time series of principal component metrics: $METRIC_{kt}^{s_{i}}=\{metric_{k1},metric_{k2},...,metric_{kN}\}^{s_{i}}$, which represents the time series of a performance metric during stage $s_{i}$ and in node k, and the length of the time series is N. In this time series, each value corresponds to a metric value at a time point.
 %where $t\in(ST_{s_{i}}, ET_{s_{i}})$, here, $ST_{s_{i}}$ and $ET_{s_{i}}$ are the start time and end time of stage $s_{i}$. Then we try to use two transformation methods for time series' data reduction.

 We use $METRIC_{kt}^{s_{i}}=\{metric_{k1},metric_{k2},...,metric_{kN}\}^{s_{i}}$ to represents the time series of principal component metrics during stage $s_{i}$  and on node $k$, and the length of the time span is $N$. Obviously, where there is a time point, there is a metric value. To accomplish the data set reduction for outlier detection, we introduce two different methods of time-frequency transformation for input data, and  the details of both two transformations are described as follows:

(1)Mean Value.
Average value comparison of the performance metric on different nodes is a typical approach for time series transformation. If there are substantial differences in average value of one performance metric between certain nodes and other nodes, then we believe that this performance metric is a potential key metric, the calculation method is shown in Formula~\ref{mean}.
%One method is comparing the average values of one performance metric on different nodes.  If the average value of a performance metric on some nodes is significantly larger than or smaller than all other nodes, this performance metric can be a potential critical performance metrics.
\begin{equation} \label{mean}
%mean(METRIC_{kt}^{s_{i}})= \frac{\sum_{t=ST_{s_{i}}}^{ET_{s_{i}}} metric_{kt}}{N}
mean(METRIC_{kt}^{s_{i}})= \frac{\sum_{t=1}^{N} metric_{kt}}{N}
\end{equation}
(2)Fast Fourier Transform.
 Phase the phase difference between sequences in time domain can result poorly when doing similarity comparison of original signal on time domain, thus transforming original data from time domain to frequency domain is an ideal try to eliminate problems of phase difference. Fast Fourier Transform (FFT) is often utilized to transform original data from time-space domain to frequency domain and vice versa, and is an efficient method for a sequence to compute its Discrete Fourier Transform (DFT). Formula\ref{FFT} is a DFT equation, where $A_r$ is the r-th coefficient of the DFT, and $metric_{kt}$ denotes the t-th sample of the time series which consists of $T$ samples, and $\iota =\sqrt{-1}$~\cite{fft}. More over, an FFT rapidly computes such transformations by factorizing the DFT matrix into a product of sparse (mostly zero) factors. As a result, FFT manages to reduce the complexity of computing the DFT from $O(n^2)$, which arises if one simply applies the definition of DFT, to $O(n\log n)$, where n is the data size.
%The Fast Fourier Transform(FFT) is used to transform the measured data from time domain to frequency domain. This preprocessing of data can benefit the comparison between sequences. Phase difference between sequences in time domain can result poorly when doing similarity comparison between data points. Transforming data to frequency domain can eliminate the phase difference problem.
%Fourier analysis converts a signal from its original domain (often time or space) to a representation in the frequency domain and vice versa. The fast Fourier transform (FFT) is a method for efficiently computing the discrete Fourier transform (DFT) of a time series (discrete data samples), or its inverse.
%The DFT is defined by the formula \ref{FFT}:
\begin{equation} \label{FFT}
A_{r}=\sum_{t=1}^{N} metric_{kt}e^{\frac{-2 \pi\iota rt}{N}} \qquad  r=1,2,....N
\end{equation}
%where $A_r$ is the r-th coefficient of the DFT, and $metric_{kt}$ denotes the $t-th$ sample of the time series which consists of T samples, and $\iota =\sqrt{-1}$~\cite{fft}.

By using the above two transformations, for all slaves of a stage, we get a reduction data set $meanSet(metric^{s_{i}})$ or $fftSet(metric^{s_{i}})$, which is used for outlier detection. In subsequent experiments, we will further compare the detection  results of using the two transformations.
\begin{algorithm}
\renewcommand{\algorithmicrequire}{ \textbf{Input:}} %Use Input in the format of Algorithm
\renewcommand{\algorithmicensure}{ \textbf{Output:}} %UseOutput in the format of Algorithm
\caption{Time Series Transformation}
\label{Transformation}
\begin{algorithmic}[1]
\REQUIRE ~~\\ %算法的输入参数：Input
Time series of principal component metrics $METRIC_{kt}^{s_{i}}$
\ENSURE ~~\\ %算法的输出：Output
Reduction data set: \\
 $meanSet(metric^{s_{i}})$ or $fftSet(metric^{s_{i}})$
  \FOR{$k=1$;$k<p$;$k++$}
   \IF {Calculate $mean(METRIC_{kt}^{s_{i}})$}
    \STATE Output $meanSet(metric^{s_{i}})$
    \ENDIF
    \IF {Calculate $fft(METRIC_{kt}^{s_{i}})$}
    \STATE Output $fftSet(metric^{s_{i}})$
   \ENDIF
   \ENDFOR
\end{algorithmic}
\end{algorithm}
\paragraph{Normalization}
Different performance metrics in a cluster normally have varied sizes and units.
%A common problem is having some features with a large range of values, while others having rather small ones. And among the selected performance metrics, different metrics may have different units,
 For instance, the units of $cpu\_usage$ and $mem\_usage$ are percentage(\%), and its value is between zero and one.
 However, the units of $diskR\_band$ and $netS\_band$  may be MB/s, which is so different with the percentage. To adjust metrics measured at the stage on different scales to a common scale notionally, normalization~\cite{Normalization} is applied into the data preprocessing, with the help of that, it would be more normal to process the data with consistent statistical properties.

 %In this case, the larger metric becomes dominant. To minimize this problem, data normalization~\cite{} is performed on the feature matrix in order to make it have consistent statistical properties.

In this section, we use the linear Min-Max Normalization to convert the original metrics into values range from 0 to 1.
%we use Min-Max Normalization,  which is a linear transformation for original data and forcing the ranges to be in [0,1].
%The conversion function is as follows formula \ref{Normalization}:
Formula~\ref{Normalization} is the transformation expression,  $y$ is a sample in $meanSet(metric^{s_{i}})$ or $fftSet(metric^{s_{i}})$, and $max$ is the maximum value of the samples, $min$ is the maximum value of the samples. However, the disadvantage of this normalization method is that max and min might be redefined when inputting the extra new data.
\begin{equation} \label{Normalization}
y^*=\frac{y-min}{max-min}
\end{equation}

\paragraph{Outlier detection}

\begin{algorithm}[htp]
\renewcommand{\algorithmicrequire}{ \textbf{Input:}} %Use Input in the format of Algorithm
\renewcommand{\algorithmicensure}{ \textbf{Output:}} %UseOutput in the format of Algorithm
\caption{An Outlier Detection Algorithm Based on Distance and Magnitude}
\label{Outlier}
\begin{algorithmic}[1]
\REQUIRE ~~\\ %算法的输入参数：Input
Normalized $meanSet(metric^{s_{i}})$ or $fftSet(metric^{s_{i}})$
\ENSURE ~~\\ %算法的输出：Output
Outlier metrics
 \IF{$min-max<=-2\log(10)$}
    \STATE $\log(10)$ conversion for input data set
    \STATE Call the magnitude-based outlier algorithm:
    \label{ code:outlier:magnitude }%对此行的标记，方便在文中引用算法的某个步骤
     \STATE (1)Find the center of mass(median);
     \STATE (2)Calculate the distance $dis$ from each point to the center of mass
         \IF{$dis>avg(dis)$}
          \STATE  Add the point into the suspicion group $SuspG$
         \ENDIF
    \STATE (3)Compute the distance $dis(SuspG)$ from the point in  $SuspG$ to the center of mass;
      \IF{ ($dis(SuspG)-avg(dis))>variance$}
       \STATE This point in  $SuspG$ is counted as outlier.
       \ENDIF
    \ELSE
    \STATE Call the distance-based outlier algorithm:
    \STATE (1)Select the maximum and minimum values for the current point in class A and B;
    \STATE (2)Calculate the distance $dis(A)$  and $dis(B)$ from each point to the two current points;
      \IF { $dis(A) < Th\_knn$}
        \STATE Assign the point to class A;
      \ELSE
        \STATE Assign the point to class B;
      \ENDIF
      \IF {$Num(A)< Num(B)$}
        \STATE  (3)Compute the distance $dis(a,B)$ from the point in A to the class B (the representative point of class B);
          \IF {$dis(a,B) > dmin$}
          \STATE This point a  is counted as outlier.
          \ENDIF
      \ELSE
          \STATE  (3)Compute the distance $dis(b,A)$ from the point in B to the class A (the representative point of class A);
           \IF {$dis(b,A) >= dmin$}
           \STATE This point b is counted as outlier.
           \ENDIF
       \ENDIF
   \ENDIF
  % \RETURN $E_n$; %算法的返回值
\end{algorithmic}
\end{algorithm}
In statistics, an outlier is an observation point that is distant from other observations~\cite{Outlier-wiki}. %In this section, we use the unsupervised outlier detection method to distinguish the metrics which do not conform to an expected pattern or other metrics in a dataset. After abstracting the metrics data, we propose an  outlier detection algorithm based on distance and magnitude.
In this section, we propose an unsupervised method combing with distance and magnitude algorithm for outlier detection to distinguish the metrics that do not belong any expected pattern in the dataset or show certain similarities with other metrics.
%Unsupervised anomaly detection techniques detect anomalies in an unlabeled test data set under the assumption that the majority of the instances in the data set are normal by looking for instances that seem to fit least to the remainder of the data set.

Our distance-based outlier model borrow ideas from the distribution-based approaches, it is also suitable for situations where the observed distribution does not fit any standard distribution~\cite{db-outlier}.
%Here, we utilize $pct$ and $dmin$ to determine the distance-based outlier, referred to as $DB(pct, dmin)$.
%The distance based notion of outliers unifies distribution based approaches.
Specifically, an object $o$ in a dataset $D$ is an $DB(pct, dmin)$-outlier, if at least a fraction $pct$ of all data objects in $D$ lies greater than distance $dmin$ from $o$.
We use the term $DB(pct, dmin)$-outlier as shorthand notation
for a Distance-Based outlier (detected using parameters $pct$ and $dmin$).
%And it is suitable for situations where the observed distribution does not fit any standard distribution~\cite{db-outlier}.
Of course, the choice of parameters $pct$ and $dmin$, and validity checking (i.e., deciding whether each $DB(pct, dmin)$-outlier is a real outlier), requires expert.

Even though our distance-based outlier algorithm with appropriate parameter settings is able to detect most of outliers in the dataset, some outliers could still be missed. For instance, the normalized mean value of $cpu\_usage$ on each node is [ hw073: 0.006838, hw106: 0.15604399, hw114: 0.17810599 ] respectively, however, there would exist no outlier as setting $dmin$ equal to 0.5 and $pct$ equal to 1. Actually, we could consider 0.006838 as an outlier value here. To make our outlier detection model still work in this case, we apply logarithm method (e.g., $\log(10)$) in the beginning to obtain data’s order of magnitude by transforming the original data, for example, the order of magnitude on several nodes  [ hw073: -2, hw106: 0, hw114: 0 ] shows significant disparity, we can suppose hw073 is a potential outlier, then the remaining two nodes would be analyzed through distance-based detection algorithm.

Algorithm~\ref{Outlier} is the detailed pseudo-code, and describe the outlier detection algorithm based on distance and magnitude.
 %the default value of $pct$  is 1, and $dmin$ is adjustable.
%The pseudo-code can be see from Algorithm~\ref{Outlier}.
In this algorithm, we predefine the parameters $pct$ as a default value 1, and  $dmin$ is adjustable.
 %can be adjusted by users.
In addition, we use two methods to calculate the representative point of class A or B, one  method is computing the max/min value of larger class, the other method is computing the median value of larger class.
%In the subsequent experiments, we will compare the outlier detection results when using different  representative points of larger class and different $dmin$.
In the subsequent experiments, we will compare the results of outlier detection by the max/min value method and the median value method with different $dmin$ for larger class.

\section{HybridTune Implementation}\label{tool}

\begin{figure*}
 \centering
 \includegraphics[width=7in,height=1.9in]{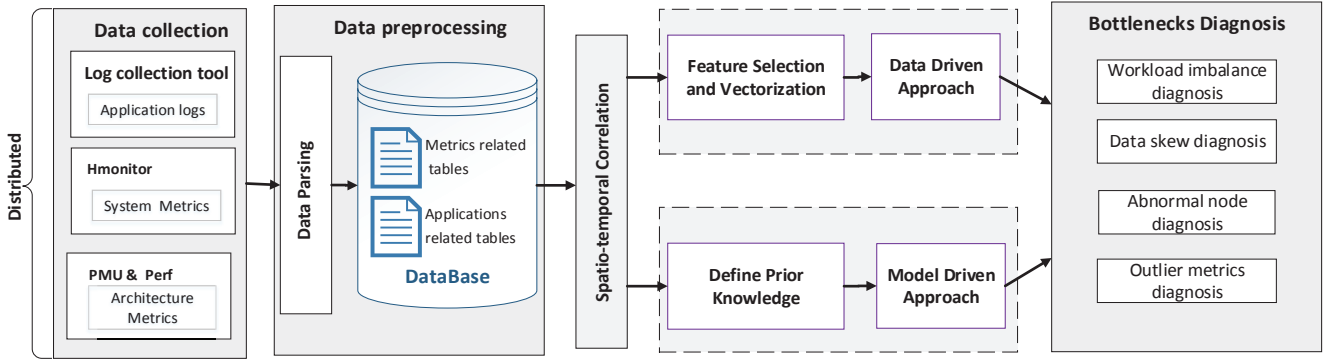}
  \caption{The workflow of HybridTune.}\label{processing-flow}
\end{figure*}

Based on our general performance diagnosis approach, we have implemented HybridTune, a scalable, lightweight, model-based and data-driven performance diagnosis tools utilizing spatio-temporal correlation.
In this section, we describe the implementation of HybridTune, and the workflow of HybridTune is shown in Fig.~\ref{processing-flow}.

\subsection{Data Collection}\label{collection}

We use the data collector of BDtune~\cite{BDTune} to gather the performance information and application logs from the software stack of Big Data systems at different levels.
%, and then sending the tracing data of different levels to the data aggregator.
Specifically, the data collector can collect architecture-level metrics, system-level metrics and application logs.
  Hardware Performance Monitoring Unit (PMU) and Perf~\cite{perf} are used for data sampling of architecture-level metrics in data collector, and metrics consists of instruction ratio, instructions per cycle (IPC), cache miss, translation lookaside buffer (TLB) miss, etc.
  Then, we use Hmonitor~\cite{BDTune} to collect raw data from the filesystem /proc, which provides key parameters (e.g., CPU usage, memory access, disk I/O bandwidth, network bandwidth) of system performance. Furthermore, we use log collection tools to collect the application logs (e.g., history job logs of Spark and Hadoop). Table~\ref{raw_metric} lists the detailed collecting information.

  \begin{table}[htbp]
\renewcommand\arraystretch{1.5}
  \caption{The format of Application Logs, Raw Architecture-level Metrics and System-level Metrics.}\label{raw_metric}
\center
\begin{tabular}{|p{8.1cm}|}
 \hline
   \textbf{\emph{The format of Application history job logs}}: \\
  \{"Event" : "SparkListenerTaskEnd" , "Stage ID" : 0, "Stage Attempt ID" :0,
   "Task Type" :"ShuffleMapTask","Task End Reason" :\{"Reason":"Success"\},"Task Info":\{"Task ID":2,"Index":2,"Attempt":0,"Launch Time" :1456896044081,"Executor ID":"0","Host":"hw073","Locality":
   "PROCESS\underline{ }LOCAL","Speculative":false,"Getting Result Time":0,"Finish Time":1456896045955,"Failed":false,
   "Accumulables":[\{"ID":1,"Name":"peakExecutionMemory","Update":"920
   ","Value":"920","Internal":true\}]\},"Task Metrics":\{"Host Name":"hw073","Executor Deserialize Time":1548,"Executor Run Time":147,"Result Size":1094,"JVM GC Time":0,"Result Serialization Time":21,"Memory Bytes Spilled":0,"Disk Bytes Spilled":0,"Shuffle Write Metrics":\{"Shuffle Bytes Written":26,"Shuffle Write Time":4201900,"Shuffle Records Written":1\}\}\} \\
  \hline
   \textbf{\emph{The format of Raw Architecture-Level Metrics}}: \\
   timestamp cycle ins L2\underline{ }miss L2\underline{ }refe L3\underline{ }miss L3\underline{ }refe DTLB\underline{ }miss ITLB\underline{ }miss L1I\underline{ }miss L1I\underline{ }hit MLP MUL\underline{ }ins DIV\underline{ }ins FP\underline{ }ins LOAD\underline{ }ins STORE\underline{ }ins BR\underline{ }ins BR\underline{ }miss unc\underline{ }read unc\underline{ }write  \\

   \hline
   \textbf{\emph{The format of Raw System-Level Metrics}}:\\
   timestamp  usr   nice    sys     idle    iowait  irq   softirq  intr   ctx     procs   running blocked   mem\underline{ }total free    buffers cached  swap\underline{ }cached   active  inactive        swap\underline{}total      swap\underline{ }free       pgin    pgout   pgfault pgmajfault    active\underline{ }conn     passive\underline{ }conn    rbytes  rpackets        rerrs   rdrop   sbytes  spackets        serrs   sdrop   read    read\underline{ }merged     read\underline{ }sectors    read\underline{ }time       write   write\underline{ }merged    write\underline{ }sectors   write\underline{ }time      progress\underline{ }io   io\underline{ }time io\underline{ }time\underline{ }weighted    \\
  \hline
  \end{tabular}
\end{table}

\subsection{Data Preprocessing}\label{preprocessing}

It is so important for performance analysis to efficient data preprocessing, since log files and performance data with non-uniform formats are generally collected from different nodes. Therefore, we  parse the performance data and application logs, and then unify the data format, preprocess the raw data and load the data into our MySQL database.

\subsubsection{Data Parsing}\label{Parsing}

In order to deal with different log formats of applications,we establish various log parsing templates compatible with different applications’ logs. In our implementation, we collected the history job logs of Hadoop and Spark, which are json formats and record various information about jobs' run. Then we parse and
extract some useful application data from these history job logs, for example:
\begin{itemize}
 \item Runtime information:  The submission time, completion time and runtime of jobs, stages and tasks.
\item Dataflow information: The data flow information between nodes in each stage of jobs, includes the reading and writing data, reading and writing time, input and output data of tasks, and so on.
\item Application configuration information: The configuration parameter information of Hadoop/Spark, etc,.
\item Job Runtime Parameters: Job-level parameters and task-level parameters, for example, the "counters" information of Hadoop and the "task\_metrics" information of Spark.
\end{itemize}

Simultaneously, we use the collected raw metrics to calculate the selected performance metrics which are shown in Table~\ref{performance_metric}.
For different performance metrics, there are different calculation methods, such as $cpu\_usage$ can be calculated by these metrics $usr$, $nice$, $sys$, $idle$, $iowait$, $irq$  and $softirq$~\cite{CPU_time}. In this section, the calculation formulas are not described in detail.

\begin{table}[htbp]
\renewcommand\arraystretch{1.5}
  \caption{The Selected Performance Metrics.}\label{performance_metric}
\center
\begin{tabular}{|p{1.4cm}|p{1.5cm}|p{4.7cm}|}
  \hline
   Layer & Metrics & Description \\     \hline
   \multirow{8}{1.5cm}{System level}   & cpu\_usage & CPU utilizations \\
       & mem\_usage   &  Memory usage \\
        & ioWaitRatio &   Percentage of CPU time spent by IO wait  \\
        & weighted\_io  & Average weighted disk io time    \\
        & diskR\_band  &  Disk Read Bandwidth   \\
       & diskW\_band &  Disk Write Bandwidth   \\
       & netS\_band  &  Network Send Bandwidth \\
       &  netR\_band &  Network Receive Bandwidth \\ \hline
    \multirow{12}{1.5cm}{Architecture level} & IPC & Instructions Per Cycle \\
    & L2\_MPKI   &  Misses Per Kilo Instructions of L2 Cache \\
       & L3\_MPKI   &  Misses Per Kilo Instructions of L3 Cache \\
      &L1I\_MPKI   &  Misses Per Kilo Instructions of L1I Cache \\
     &ITLB\_MPKI   &  Misses Per Kilo Instructions of ITLB  \\
     &DTLB\_MPKI   & Misses Per Kilo Instructions of DTLB  \\
      &MUL\_Ratio   &  MUL operation' percentage  \\
       &DIV\_Ratio   &  DIV operation' percentage  \\
        &FP\_Ratio   & Floating point operations' percentage \\
     &LOAD\_Ratio   & Ratio of LOAD Operation  \\
       &STORE\_Ratio   & Store operations' percentage   \\
      &BR\_Ratio   & Branch operations' percentage   \\
    \hline
  \end{tabular}
\end{table}

\subsubsection{Data Storage}\label{aggregator}

%It is so important for performance analysis to efficient data manipulation, since log files and performance data with non-uniform formats are generally collected from different nodes.Therefore, we build the data aggregator to classify, group and parse the performance data and logs gathered by data collector, and then unify the data format, preprocess the raw data and load the data into our MySQL database.

 In addition, we design a tagging mechanism and propose an incremental table approach to match the scalability need of date aggregation and storage.
 Specifically, we set corresponding labels for tables of different applications, for instance, if label $Type\_Flag = 0$, then the tables represent the parsed Hadoop log contents, if $Type\_Flag = 1$, we know that tables store the parsed Spark logs.
 Moreover, we provide public table interfaces and unique table interfaces for different application logs, because the contents of application logs are not always the same. For example, both Hadoop and Spark consist of public table interfaces like $app$ table, $job$ table, $stage$ table and $task$ table. The unique table interfaces for Hadoop are $task\_attempt$ table and $counters$ table, unlike Hadoop, Spark’s unique table interfaces are $rdd$ table and $task\_metrics$ table, and so on.
 When collecting Storm logs and storing parsed data into MySQL, the data preprocessor needs to adjust its log parsing template that only for Storm applications, then it creates unique tables of Storm, parsing data, preprocessing raw data, loading data into existed public tables and its unique tables, and setting $Type\_Flag$.

% Afterwards, the loaded raw data need to be preprocessed. There are mainly three parts of data preprocessing: invalid and missing data handling, performance metrics computing at system level and architecture level, and statistical information generating.

\subsection{Data \& Model Driven Performance Diagnosis}

After data preprocessing, performance diagnosis model would undertake the specific analysis and diagnosis works. Our performance diagnosis model is equipped with a plug-in mechanism, which enables the analysis engine to
%be modified according to the certain situation for
adapt different application occasions and different diagnosis methods. Among these plug-in mechanisms, some are universal (e.g., statistical analysis of performance metrics), while some are application-specific (e.g., critical path computing of different jobs).

Specifically, the workflow of performance diagnosis method comprises three main step: (1) spatio-temporal correlation, (2) feature selection and vectorization and using the data driven approach / defining prior knowledge and using the model driven approach,  (3) bottlenecks diagnosis. The details  are described in section~\ref{diagnosis-method}.

\section{Evaluations}\label{Evaluations}

\subsection{Experiment Settings}

% The Hadoop cluster used in our experiment consists of 1 master and up to 6 slaves, and the Spark cluster is deployed upon the Hadoop Yarn framework. Specifically, the ResourceManager and NameNode of Hadoop and the Master of Spark run on the master node, the NodeManager and DataNode of Hadoop and the Workers of Spark run on the slaves. In order to ensure clock synchronization across nodes, we use NTP service in the cluster. The detailed server configurations of the cluster are shown in Table \ref{servers}.

 %In addition, since we do not evaluate the impacts of configurations on performance, we use homogeneous clusters with the same machine configurations, for example, the hardware thread number and memory capacity are same in every worker.%, the data size of each task processes is about 128MB.

The Hadoop cluster used in our experiment consists of one master machine and six slave machines, the Resource Manager and Name Node modules are deployed on the master node, the Node Manager and Data Node only run on the slave node. Furthermore, we deploy our Spark cluster on the Hadoop Yarn framework, the Master module runs on the master machine, and the Workers executes on slave machines. In our cluster, we use NTP (Network Time Protocol) service to ensure clock synchronization across nodes, Table~\ref{servers} details the server configurations of our cluster.   In addition, the evaluations about impacts of configurations on system performance are not included in this paper, thus our machines in the cluster are  homogeneous  machines with the same machine configurations and cluster configuration parameters.
\begin{table}[htbp]
\caption{Server configurations}\label{servers}
\begin{tabular}{| p{1.4cm} | p{6cm} |}
  \hline
     Processor &  Intel(R) Xeon(R) CPU E5645@2.40GHz  \\ \hline
     Disk   & 8 Seagate Constellation ES (SATA 6Gb/s)- ST1000NM0011 [1.00 TB] \\ \hline
     Memory  & 32GB  per server \\ \hline
     Network &  Broadcom Corporation NetXtreme II BCM5709 Gigabit Ethernet (rev 20) \\ \hline
     Kernel & Linux Kernel 3.11.10\\
  \hline
\end{tabular}
\end{table}

\subsection{Anomaly Simulation}

%Furthermore, in order to evaluate the effect of automatic diagnostic tool, we use the following methods to simulate anomalies:
To further determine whether or not workload imbalances, straggler nodes, data skew and abnormal machine states exist, and evaluate the effectiveness of our automatic diagnosis tool for performance bottleneck detection, we decide to simulate anomalies by the following methods:

(1) Reduce the computing power of some nodes. %for example, turning off some cores on a multi-core machine. However, when turning off cores, the core 0 can not be turned off and the closed cores can not be too much, otherwise it may cause entire system to crash.
For example, you can attempt to disable a core in a multi-core machine. However, this does not work for a few certain cores, like the core 0. And disabling too many cores would result in a system crash.

(2) %Make uneven disk storage on some nodes, for example, filling the disks on some nodes.
Make disk storage imbalance. E.g., filling the space of disks on some nodes with data.

(3) %Mixed Running interference procedures on some nodes, for example, we use the Linux stress test tool \raisebox{0.5mm}{------} Stress~\cite{stress} to impose pressures at CPU, memory, IO and disk, etc.
Mix interference workloads. In our experiment, Linux stress testing tool \raisebox{0.5mm}{------} Stress~\cite{stress} is utilized to impose extra load on CPU, memory, IO and disk.

(4) %Use a cache flusher to load certain volume of data according to the size of the last level cache to insert cache anomalies.
Cache flusher adjustment. We use a cache flusher to   load certain volume of data according to sizes of the last-level cache, for inserting cache anomalies.

%Through the above methods to simulate anomalies, we can further investigate whether there are workload imbalances, straggler nodes, data skew or machine abnormal states, and find out and diagnose the causes for performance bottlenecks.

\subsection{Anomaly detection Evaluation Results}

%In the experiments, we run 90 programs, which including 342 stages. Among them, we run the workloads include Wordcount, Grep, Sort, Kmeans in BigDataBench, FPGrowgh and PrefixSpan in Spark Mllib~\cite{mllib}.
In the section, over 90 programs on 342 stages are tested and executed. We also try a number of workload executions, e.g., Wordcount, Grep, Sort, K-means on BigDataBench, FPGrowgh and PrefixSpan on Spark MLlib~\cite{mllib}. Due to the bottlenecks at application level (such as workload imbalance and data skew) may be related to the users' subjective view even more, so we plan to use the artificially setting threshold, for example, we can set BC=0.1, $Th\_UB$=0.6 and $Th\_size$=1.5~\cite{BDTune}. Here, we mainly evaluate the threshold selection of abnormal node and outlier metrics, as well as the effect of outlier detection.

%\begin{figure*}
 %\centering
 %\includegraphics[width=6in,height=2in]{accuracy.eps}
 % \caption{The effectiveness of outlier detection.}\label{accuracy}
%\end{figure*}

\subsubsection{Abnormal node determine}

\begin{figure}
 \centering
 \includegraphics[width=2.6in,height=1.5in]{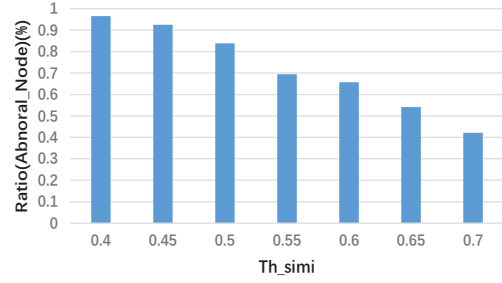}
  \caption{The results of $Ratio(Abnormal\_Node)$.}\label{simi}
\end{figure}

To determine whether  or not node $k$ is an abnormal node, we compare  $simi(\overrightarrow{v_{k}},\overrightarrow{v_{others}})$ (formula~\ref{AvgSimilarity} in section~\ref{abnormal_node_detection}) with the predefined threshold $Th\_simi$. If the similarity value is smaller than the threshold, then the node is judged as an abnormal node.% So it involves how to determine the value of the threshold $Th\_simi$.
As for how to set the size of threshold $Th\_simi$, we utilize formula~\ref{measure-simi} to measure the proportion of real abnormal nodes in the detected abnormal nodes detected just by predefined threshold $Th\_simi$, and the results are shown in Figure~\ref{simi}.
%Figure~\ref{simi} shows that detailed results. %shift over the size of threshold.
%In our experiments, we use formula \ref{measure-simi} to indicate the proportion of real abnormal nodes in the detected abnormal nodes, when predefining different threshold $Th\_simi$, and the results are shown in figure~\ref{simi}.
\begin{equation} \label{measure-simi}
\begin{split}
&Ratio(Abnormal\_Node)= \frac{\# \ of \ abnormal \ nodes}{\# \ of \  detected \ abnormal \  nodes }
\end{split}
\end{equation}

%\begin{figure}
% \centering
% \includegraphics[width=2.5in,height=1.5in]{simi.eps}
 % \caption{The proportion of real abnormal nodes in the detected abnormal nodes.}\label{simi}
%\end{figure}

 \subsubsection{Selection of principal components}

%Since the dimension of principal component metircs will affect the results of outlier metircs detection, and the cumulative contribution rate is usually used for determining an appropreate dimension of the eigenspace, so we set different $CCRate_d$ and select an appropriate value in our experiments.

The results of outlier metrics detection are often affected by the dimension of principal component metrics. To determine an appropriate dimension for the eigenspace, we decide to use the cumulative contribution rate  $CCRate_d$  to help selecting principal components. First we de?ne $PC_i$ as the $i$-th principal component, then we calculate the ratio of a certain performance metric $M_x$ to principal component $PC_i$, as shown in formula~\ref{measure-pca1}. We discover that the principal component $PC_i$ always belongs to several particular performance metrics.
%In detail, we define $PC_i$ to indicate the $i$-th principal component, and we find that a principal component $PC_i$  is always a specific performance metric. We use formula~\ref{measure-pca1} to calculate the ratio of performance metric $M_x$ is principal component $PC_i$.
\begin{equation} \label{measure-pca1}
\begin{split}
&Ratio(M_x,PC_i)= \frac{\# \ of \ M_x \ that \ is  \ PC_i}{\# \ of \ PC_i}
\end{split}
\end{equation}

\begin{figure*}
 \centering
 \includegraphics[width=7in,height=2.5in]{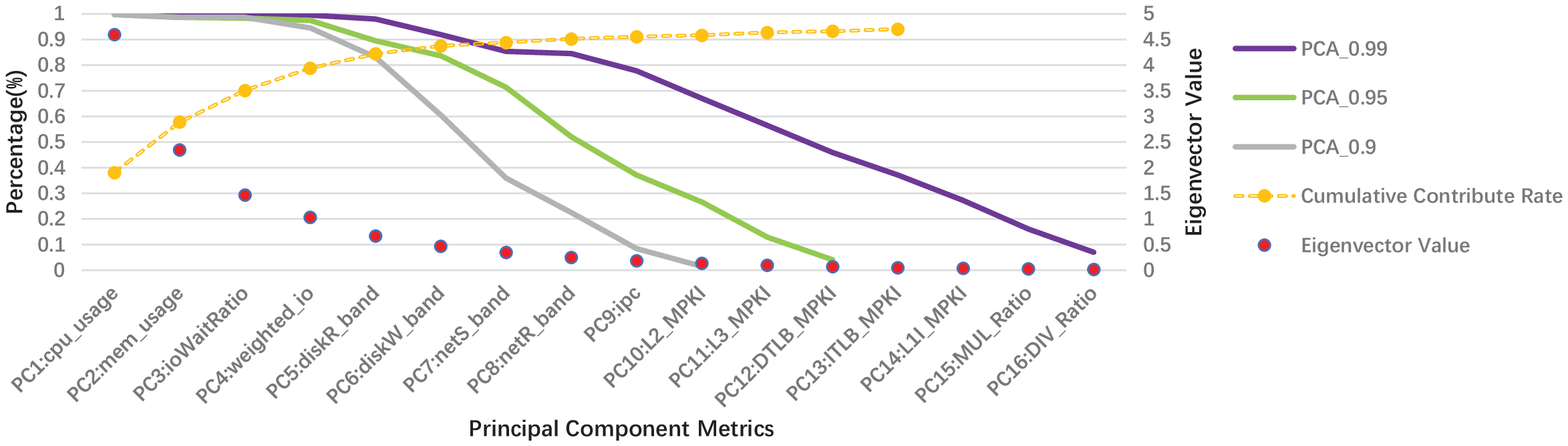}
  \caption{The principal component results by PCA analysis.}\label{pca}
\end{figure*}

We can see various average eigenvector values, the cumulative contribute rate of performance metrics and $Ratio(PC_i,PCA)$  from Figure~\ref{pca}. $Ratio(PC_i,PCA)$ here refers to the proportion of a principal component $PC_i$  to the total number of PCA, which is illustrated in formula~\ref{measure-pca2}.
Figure~\ref{pca} also show the different results of principal components selection while setting the cumulative contribution rate as 0.9, 0.95 and 0.99. If we choose 0.9 as the cumulative contribution rate, the principal components would involve metrics from PC1 to PC10, and among them, only PC9 metric and PC10 metric belongs to architecture-level metrics. However, if our cumulative contribution rate equals to 0.95, a few system-level metrics and architecture-level metrics are selected as the principal components. Furthermore, when the cumulative contribution rate is 0.99, almost all metrics would be contained in principal components, this makes the cumulative contribution rate become meaningless. Above all, we choose 0.95 as the cumulative contribution rate for our principal components selection.
%Figure~\ref{pca} shows that the average eigenvector value, cumulative contribute rate of performance metric and  $Ratio(PC_i,PCA)$ (see formula~\ref{measure-pca2}), which indicates the proportion of a principal component $PC_i$ to the total number of PCA analysis. In addition, we compared the selected principal components when setting different cumulative contribution rate \raisebox{0.5mm}{---} 0.9, 0.95 and 0.99: for example, when cumulative contribution rate is 0.9, the selected principal components are from PC1 to PC10, and only PC9 and PC10 are architecture-level metrics; when cumulative contribution rate is 0.95, the selected principal components contain system-level metrics and a part of architecture-level metrics; however, there is little significance when cumulative contribution rate is 0.9, because it will output almost all metrics. In summary, it is recommended to select 0.95 as the threshold of the cumulative contribution rate.
\begin{equation} \label{measure-pca2}
\begin{split}
&Ratio(PC_i,PCA)= \frac{\# \ of \ M_x \ which \ is  \ PC_i}{\# \ of \ PCA}
\end{split}
\end{equation}

\begin{figure*}
\begin{minipage}{0.48\linewidth}
\centering
\includegraphics[width=3.38in]{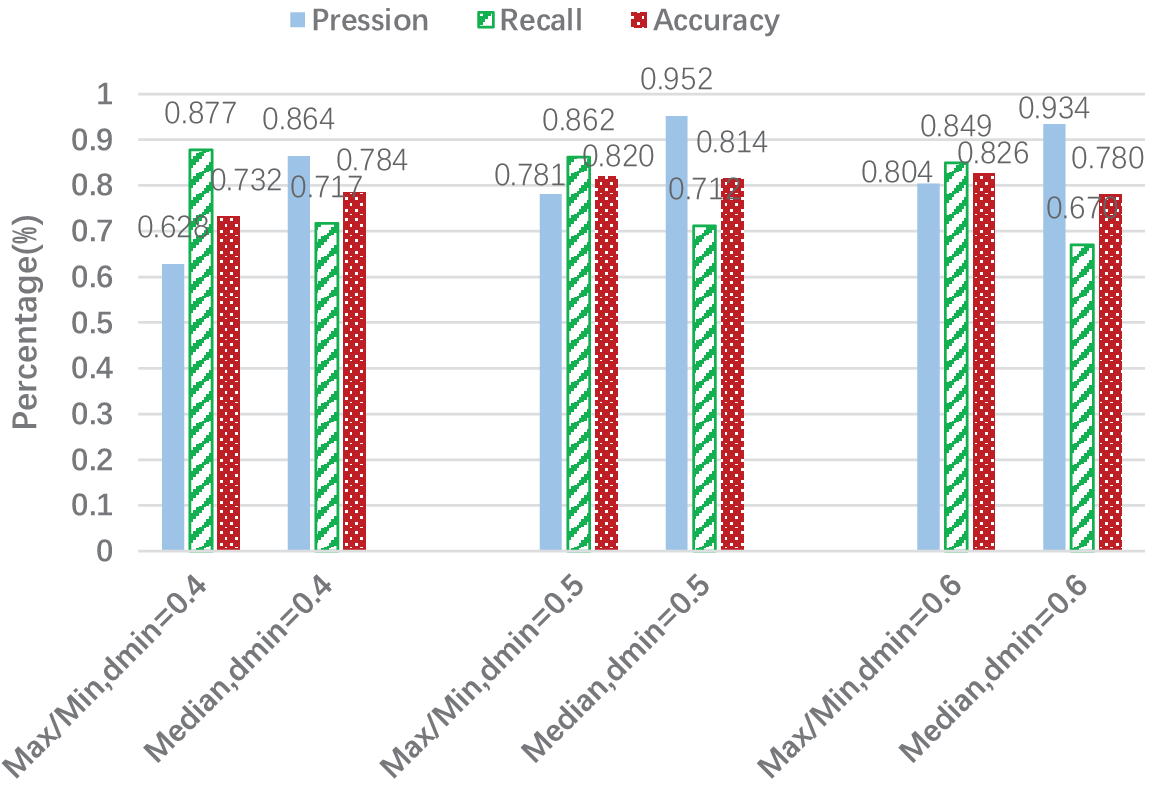}
\caption{The effectiveness of outlier detection when using Mean-Value transformation.}
\label{pression_mean}
\end{minipage}%
\begin{minipage}{0.52\linewidth}
\centering
\includegraphics[width=3.41in]{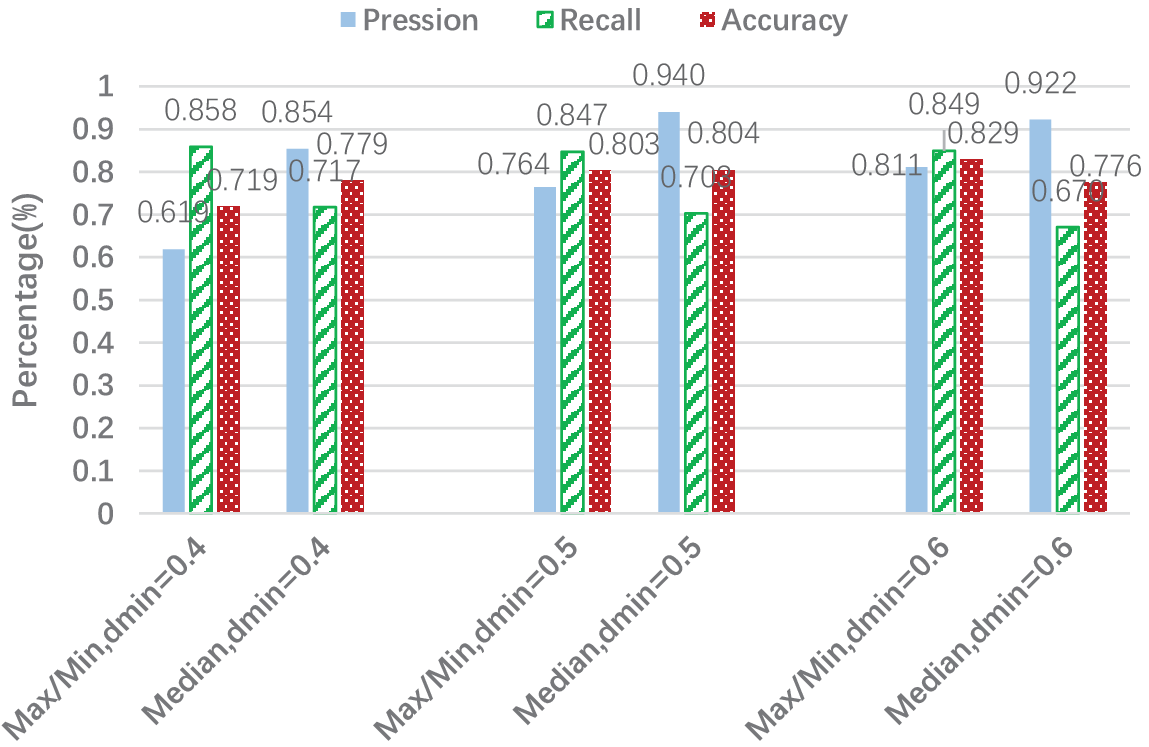}
\caption{The effectiveness of outlier detection when using FFT transformation.}
\label{pression_fft}
\end{minipage}
\end{figure*}

 \subsubsection{The effect of outlier detection }
We evaluate the effectiveness of outlier metrics detection through three indicators: Precision, Recall and Accuracy~\cite{detection-metric}. $Precision$ means the exactness of outlier detection, the higher the precision we have, the lower the rate of false alarms we get. $Recall$ indicates to the completeness of detection, the higher the recall is, the lower the false negative rate we obtain. Nevertheless, only a precision or a separate recall is incapable of evaluating the effectiveness of an anomaly detection method. Therefore, we introduce the $Accuracy$ (the harmonic mean of $Precision$ and $Recall$) to measure outlier metrics detection.
%We use three measures to evaluate the effectiveness of outlier detection~\cite{detection-metric}:
\begin{equation} \label{measure-1}
\begin{split}
&Precision= \frac{\# \ of \ successful \ detections}{\# \ of \ total \ alarms}
\end{split}
\end{equation}
\begin{equation} \label{measure-2}
\begin{split}
&Recall= \frac{\# \ of \ successful \ detections}{\# \ of \ total \ outliers}
\end{split}
\end{equation}
\begin{equation} \label{measure-3}
\begin{split}
&Accuracy= \frac{2*Precision*Recall}{Precision+Recall}
\end{split}
\end{equation}
%$Precision$ indicates the exactness of the detection, and $Recall$ indicates the completeness. Furthermore, the higher the $Precision$ is, the lower the false alarm rate is; and the higher the $Recall$ is, the lower the  false negative rate is. %1-Precision indicates the false alarm rate, and 1-Recall indicates the false negative rate. However, neither $Precision$ nor $Recall$ alone can judge the goodness of an anomaly detection method. So we introduce $Accuracy$, the harmonic mean of $Precision$ and $Recall$.

We can see the similar results of the outlier metrics detection by Mean-Value transformation and FFT transformation from figure~\ref{pression_mean} and figure~\ref{pression_fft} respectively. If we utilize the median value to represent larger class, then $Precision$  is higher than $Recall$. As is shown in the figure~\ref{pression_mean} and figure~\ref{pression_fft}, when $dmin$ equals to 0.5 or 0.6, the $Precision$ reaches more than 92\%, however $Recall$ is a slightly lower: 67\% and 70\% respectively. In addition, if using max/min value as the larger class, in contrast, $Recall$ would be higher than $Precision$, we can see the $Recall$s in figure~\ref{pression_fft} and figure~\ref{pression_mean} are both over 84\% no matter what the $dmin$ is. In other words, there are more outliers able to be detected and the false negative rate is much lower. Contrary to the $Recall$ , the $Precision$ is a little low. If the $dmin$ value is 0.4, 0.5 or 0.6, then the $Precision$ ranges from 62\% to 81\%, it means that lots of normal metrics are misjudged as outliers, thus resulting in a high false positive rate. Additionally, when max/min value represents the larger class and $dmin$ is 0.6, then the $Accuracy$ closes to 83\%. However, if $dmin$ is set to 0.5, the $Accuracy$ is about 80\% when using max/min value and median value, and setting $dmin$ to 0.4 would further decrease the $Accuracy$.

\section{Case Studies }\label{cases}
From~\cite{BDTune} we know that, the causes of performance bottleneck can be classified four categories: improper configuration, data skew, abnormal nodes and intra-node resource interference. In this section, based on our detection and diagnostic methods,
we share our experiences on tuning and diagnosing the performance of Spark and Hadoop applications, illustrate the three cases in the paper~\cite{BDTune} and an added case for Hadoop job.

\subsection{Case-1: Uneven Data Placement}

From ~\cite{BDTune} we know that, the S-WordCount job's stage \emph{spark\underline{ }stage\underline{ }app-20160630230531-0000\underline{ }0} have a straggle outlier node hw114 when  Th\_D=1.5, and  workload  imbalance when BC=0.1. In this paper, we give the automatic diagnostic results in Table~\ref{Diagnostic-1}.

 In contrast to other nodes, the priority of data locality of hw114 is "ANY" and the average similarity between hw114 and other slave nodes is 23.57\%. That comes about because, the uneven data phenomenon placement exists in hw114. We find that every task on node hw114 needs to read data from other nodes rather than the local, so that their task runtime is relatively longer. So we decide to utilize the HDFS (Hadoop Distributed File System) balancer to optimize the data distribution, then the completion time of this S-WordCount job is reduced from 218 seconds to 167 seconds, approximately 23.21\%.

In addition, we also give a  automatic diagnostic results of Hadoop's \emph{mapStage\underline{ }job\underline{ }1493084522519\underline{ }0014} in Table~\ref{Diagnostic-1-2}. We can know that this map stage of Hadoop has no obviously straggle outlier node, but there exist workload imbalances. In fact, we check the task assignments of hw106, hw114, hw062 and hw073, which are respectively 228, 159, 44 and 23. It is obvious the  assigned tasks in hw073 and hw062 are significantly less than that in the other two nodes, and the reason is that the localities of hw073 and hw062 are  "RACK\_LOCAL", while the localities of hw106 and hw114 are "NODE\_LOCAL".
  \begin{table}[tbp]
\renewcommand\arraystretch{1.5}
  \caption{The automatic diagnostic results of Spark job for case-1 .}\label{Diagnostic-1}
\center
\begin{tabular}{|p{8.2cm}|}
  \hline
  \hangafter=1
  \textbf{\emph{ Stage id}}: \emph{spark\underline{ }stage\underline{ }app-20160630230531-0000\underline{ }0} \\
   \textbf{\emph{ Detected straggle outlier node}}: hw114 \\
    \textbf{\emph{ Detected workload imbalance}}: hw114\\
    \textbf{\emph{---  Data skew diagnosis}}: \\
      \textbf{\emph{~~~~~ Skew data size：}}: Null \\
    \textbf{\emph{~~~~~ Uneven data placement：}}: hw114 [ANY:0.06875] \\
   \textbf{\emph{---  Abnormal node diagnosis}}: \\
   \textbf{\emph{~~~~~ Similarity analysis：}}: Similarity (['hw089', 'hw062', 'hw073',\\
    ~~~~~ 'hw103', 'hw114', 'hw106'], other nodes): [0.8048, 0.7838, \\
    ~~~~~ 0.8242, 0.7870, 0.2359, 0.8171]\\
    \textbf{\emph{~~~~~  Detected abnormal node}}: hw114  \\
     \textbf{\emph{---  Outlier metrics diagnosis}}: \\
     \textbf{\emph{ ~~~~~ Mode}}: [Mean-Value,median,$CCRate_d$=0.95,dmin=0.5]:\\
      hw114:(mem\_usage,ioWaitRatio,diskR\_band,netS\_band,netR\_band) \\
  \hline
  \end{tabular}
\end{table}

 \begin{table}[tbp]
\renewcommand\arraystretch{1.5}
  \caption{The automatic diagnostic results of Hadoop job for case-1.}\label{Diagnostic-1-2}
\center
\begin{tabular}{|p{8.2cm}|}
  \hline
  \hangafter=1
  \textbf{\emph{ Stage id}}: \emph{mapStage\underline{ }job\underline{ }1493084522519\underline{ }0014 } \\
   \textbf{\emph{ Detected straggle outlier node}}: Null \\
    \textbf{\emph{ Detected workload imbalance}}: hw106, hw073, hw062, hw114\\
    \textbf{\emph{---  Data skew diagnosis}}: \\
      \textbf{\emph{~~~~~ Skew data size：}}: Null \\
    \textbf{\emph{~~~~~ Uneven data placement：}}: hw073 [RACK\_LOCAL:0.09469], \\
     ~~~~~  hw062 [RACK\_LOCAL:0.00379] \\
   \textbf{\emph{---  Abnormal node diagnosis}}: \\
   \textbf{\emph{~~~~~ Similarity analysis：}}: Similarity (['hw062', 'hw073','hw114',\\
    ~~~~~ 'hw106'], other nodes): [0.8571, 0.8143,0.8784, 0.8807] \\
    \textbf{\emph{~~~~~  Detected abnormal node}}: Null  \\
     \textbf{\emph{---  Outlier metrics diagnosis}}: \\
     \textbf{\emph{ ~~~~~ Mode}}: [Mean-Value,median,$CCRate_d$=0.95,dmin=0.5]: Null\\
  \hline
  \end{tabular}
\end{table}

\subsection{Case-2: Abnormal node}
  \begin{table}[tbp]
\renewcommand\arraystretch{1.5}
  \caption{The automatic diagnostic results for case-2.}\label{Diagnostic-2}
\center
\begin{tabular}{|p{8.2cm}|}
  \hline
  \hangafter=1
  \textbf{\emph{ Stage id}}: \emph{spark\underline{ }stage\underline{ }app-20160719212517-0001\underline{ }2} \\
   \textbf{\emph{ Detected straggle outlier node}}: hw089 \\
    \textbf{\emph{ Detected workload imbalance}}: hw089 \\
    \textbf{\emph{---  Data skew diagnosis}}: \\
         \textbf{\emph{~~~~~ Skew data size：}}: Null \\
    \textbf{\emph{~~~~~ Uneven data placement：}}: Null \\
   \textbf{\emph{---  Abnormal node diagnosis}}: \\
   \textbf{\emph{~~~~~ Similarity analysis：}}: Similarity (['hw089', 'hw062', 'hw073',\\
    ~~~~~ 'hw103', 'hw114', 'hw106'], other nodes): [0.1198, 0.7667,0.8017 \\
    ~~~~~ 0.7774,0.7995, 0.7974]\\
    \textbf{\emph{~~~~~  Detected abnormal node}}: hw089  \\
     \textbf{\emph{---  Outlier metrics diagnosis}}: \\
      \textbf{\emph{ ~~~~~ Mode}}: [Mean-Value,median,$CCRate_d$=0.95,dmin=0.5]:\\
    ~~~~~~ hw089:(cpu\_usage, ioWaitRatio,weighted\_io) \\
  \hline
  \end{tabular}
\end{table}

 We also give the automatic diagnostic results of the case \emph{spark\underline{ }stage\underline{ }app-20160719212517-0001\underline{ }2}(reported in the paper ~\cite{BDTune}) in Table~\ref{Diagnostic-2}. From the automatic diagnostic results we know that, hw089 is a straggle outlier node and has workload imbalance, and it is a abnormal node whose similarity between others is about 11.98\%. Nevertheless, these bottlenecks problems are not mainly caused by data skew. However,  we do find some outlier metrics, and just find an abnormal metric: the average weighted\_io of hw089 calculated by Mean-value method is -4177890.23, contrary to common sense. Since the io\_time\_weighted value is 4294936240 at 2016-07-19 22:35:49, while the io\_time\_weighted value is 258900 at 2016-07-19 22:35:50. In order to diagnose the root cause, we further view the system logs, and find that the disk of hw089 has experienced a high temperate alarm and Raw Read Error Rate~\cite{BDTune}.

 We can see from case-2, it is necessary to analyze the correlation between the outlier metrics, for some outlier metrics may be caused by other metrics, such as case-2, the abnormal metric weighted\_io lead to the outlier metrics cpu\_usage and ioWaitRatio. Further more, in order to locate the root cause of  abnormal metric, the diagnosis based on system or RAS logs is also needed.

\subsection{Case-3: Intra-Node Resource Interference}

  \begin{table}[tbp]
\renewcommand\arraystretch{1.5}
  \caption{The automatic diagnostic results for case-3.}\label{Diagnostic-3}
\center
\begin{tabular}{|p{8.2cm}|}
  \hline
  \hangafter=1
  \textbf{\emph{ Stage id}}:  \emph{spark\underline{ }stage\underline{ }app-20160703145107-0001\underline{ }0} \\
   \textbf{\emph{ Detected straggle outlier node}}: hw062,hw106 \\
    \textbf{\emph{ Detected workload imbalance}}:  \\
    \textbf{\emph{---  Data Skew diagnosis}}: \\
           \textbf{\emph{~~~~~ Skew data size：}}:  Null \\
    \textbf{\emph{~~~~~ Uneven data placement：}}: Null \\
   \textbf{\emph{---  Abnormal node diagnosis}}: \\
   \textbf{\emph{~~~~~ Similarity analysis：}}: Similarity (['hw089', 'hw062', 'hw073',\\
    ~~~~~ 'hw103', 'hw114', 'hw106'], other nodes): [0.9593, 0.9255,0.9228 \\
    ~~~~~ 0.9437,0.9513, 0.9432]\\
    \textbf{\emph{~~~~~  Detected abnormal node}}: Null  \\
     \textbf{\emph{---  Outlier metrics diagnosis}}: \\
      \textbf{\emph{ ~~~~~ Mode}}: [FFT,median,$CCRate_d$=0.95,dmin=0.5]:\\
    ~~~~~~ hw062:(L3\_MPKI); ~~ hw106:(L3\_MPKI) \\
  \hline
  \end{tabular}
\end{table}
The automatic diagnostic results of the case \emph{spark\underline{ }stage\underline{ }app-20160703145107-0001\underline{ }0} (reported in the paper ~\cite{BDTune}) in Table~\ref{Diagnostic-3}. From the automatic diagnostic results we know that, there exists two straggle outlier nodes hw062 and hw106, and it not caused by data skew and abnormal node. However, the automatic diagnosis tool of BDTune find that the average L3\_MPKI in these two nodes are both larger than other nodes while the node similarity of all nodes in the cluster is 93.3\%.

\section{Related work}\label{related-work}

\textbf{Performance analysis}. %Currently, there are some performance analysis tools just for MapReduce application.
 There have been much prior studies on building tools to analyze performance for MapReduce applications.
 SONATA \cite{SONATA} propose a correlation-based performance analysis approach for full-system MapReduce optimization, it correlates different phases, tasks and resources for identify critical outliers and recommends optimization suggestions based on embedded rules, which just uses the model-based method.
 HiTune \cite{hitune} describe a dataflow-driven performance analysis approach, it reconstruct the high level, dataflow-based, distributed and dynamic execution process for each Big Data application. Mochi \cite{Mochi} is a visual, log-analysis based debugging tool correlates Hadoop’s behavior in space, time and volume, and extracts a causal, unified control and dataflow model of Hadoop across the nodes of a cluster.

 Besides the above tools used to analyze MapReduce applications, tools for other platforms are also proposed. Kay et al. \cite{Kay-NSDI15} use blocked time analysis to quantify the performance bottlenecks in Spark framework, and Microsoft use Artemis \cite{Artemis} to analyze Dryad application, which is  a plug-in mechanism which using statistical and machine learning algorithms.

\textbf{Performance anomaly detection and diagnosis}. In general,
% anomaly detection is required in performance diagnosis usually.
anomaly detection is an essential part of performance diagnosis for Big Data systems.
And anomaly detection techniques can be broadly classified into two groups: data-driven and model-based.Data-driven
methods include nearest neighbor based methods include distance-based~\cite{distance-based-outliers}, k-nearest
neighbor~\cite{knn} and local outlier factor~\cite{lof}, and kmeans clustering~\cite{Findout}. Specifically, A number of node comparison methods have been
adopted for anomaly detection in large-scale systems~\cite{Non-Parametric-Anomaly}. For example, Kahuna \cite{Kahuna} aims to diagnose performance in MapReduce systems based on the hypothesis that nodes exhibit peer-similarity under fault-free conditions, and that some faults result in peer-dissimilarity.  Ganesha \cite{Ganesha} is a black-box diagnosis technique that examines OS-level metrics to detect and diagnose faults in MapReduce systems, especially can diagnose faults that manifest asymmetrically at nodes. Eagle~\cite{Eagle} is a framework for anomaly detection at eBay, which uses density estimation and PCA algorithms for user behavior analysis. Kasick et al.~\cite{Black-box-file} developed anomaly detection mechanisms in distributed environments by comparing system metrics among nodes. Z. Lan et al.~\cite{Non-Parametric-Anomaly} present a practical and scalable anomaly detection method for large-scale systems, based on hierarchical grouping, non-parametric clustering, and two-phase majority voting.

Representative model-based techniques include rule based methods \cite{vaidya}\cite{Digging-Deeper},
support vector machine (SVM) based methods~\cite{svm-detection}, probability model~\cite{Probabilistic-diagnosis}, bayesian
network based methods~\cite{bayesian-detection}, etc. For example, Hadoop vaidya\cite{vaidya} is a rule-based performance diagnostic tool from MapReduce jobs. Although it can provise recommendations based on the analysis of runtime statistics, it cannot facilitate full-system optimization.  CloudDiag~\cite{Fine-grained-diagnosis} can efficiently pinpoint fine-grained causes of the performance problems through a black-box tracing mechanisms and
 without any domain-specific knowledge.  Mantri \cite{Mantri} is a system that monitors tasks and culls outliers based on their causes, and then delivers effective mitigation of outliers in MapReduce networks.

 Moreover, the pure data driven diagnosis approach is promising for relatively simple distributed applications, but it is very time-consuming and
difficult to be used in the complex Big Data systems. The model driven approach requires more detailed prior knowledge to achieve better accuracy, and it is also difficult to adapt for big data scale. Distinguished from the above works, HybridTune is a lightweight and extensible tool, which uses a hybrid method that combining with data driven and model driven diagnosis approach. It provides fine-grained spatio-temporal correlation analysis and different diagnosis.
 Due to the stage-based and multi-level performance data correlation,  it is also easily  extended to semi-realtime detection and can improve the time effectiveness of diagnosis.

  %propose a spatio-temporal correlation analysis approach based on stage characteristic and distribution characteristic of Big Data applications, which can associate the multi-level performance data fine-grained. On the basis of correlation data, we build some suitable datasets for different performance bottlenecks through feature selection and vectorization, and then utilize the model-based and data-driven algorithms for bottlenecks detection and diagnosis. Moreover, it is easily  extended to semi-realtime detection and can improve the  time effectiveness of diagnosis.

\section{Conclusion}\label{Conclusion}

 In this paper, we propose a spatio-temporal correlation analysis approach based on stage characteristic and distribution characteristic of Big Data applications, which can associate the multi-level performance data fine-grained. On the basis of correlation data,
 we define some rules, build some suitable datasets through feature selection and vectorization for different performance bottlenecks, such as, workload imbalance, data skew, abnormal node and outlier metrics. And then, we utilize the model-based and data-driven algorithms for bottlenecks detection and diagnosis. In addition,  we design and develop a lightweight, extensible tool HybridTune, and  validate the diagnosis effectiveness of our tool with BigDataBench on several benchmark experiments in which the outperform state-of-the-art methods. Our experiments show that the accuracy of abnormal/outlier detection we obtained reaches about 80\%. Furthermore, we report several Spark and Hadoop use cases, which are demonstrated how HybridTune supports users to carry out the performance analysis and diagnosis efficiently on the Spark and Hadoop applications. we can see that our model-based and data-driven detection and  diagnosis methods based on  spatio-temporal correlation data can  pinpoint performance bottlenecks and provide performance optimization recommendations for Big Data applications.
\ifCLASSOPTIONcompsoc
  \section*{Acknowledgments}
\else
  \section*{Acknowledgment}
\fi

\ifCLASSOPTIONcaptionsoff
  \newpage
\fi



%\bibliographystyle{IEEEtran}
%\begin{thebibliography}{1}

\bibliographystyle{plain}
\bibliography{references}

\end{document}